\documentclass[journal]{IEEEtran}

\usepackage{times}  
\usepackage{hyperref}
\usepackage[pdftex]{graphicx}

\hypersetup{pdfstartview=FitH,pdfpagelayout=SinglePage}

\begin{document}

\title{Internet Photonic Sensing:  Using Internet Fiber Optics for Vibration Measurement and Monitoring}

\author{\IEEEauthorblockN{Shreeshrita Patnaik, Paul Barford, Dante Fratta, Bill Jensen, Neal Lord, Matt Malloy, and Herb Wang}\\
University of Wisconsin-Madison}

\maketitle

\begin{abstract} In this paper, we introduce {\em Internet Photonic Sensing} (IPS), a new framework for deformation and vibration measurement and monitoring based on signals that are available from standard fiber optic communication hardware deployed in the Internet.  IPS is based on the hypothesis that atmospheric, seismic, anthropogenic and other natural activity cause vibrations in the earth that trigger detectable changes in standard optical signals that transmit data through Internet fiber.  We assume a simple system component model for optical communication hardware and identify two candidate signals that may reflect deformation and vibrations and that can be measured through standard interfaces:  Optical Signal Strength (OSS) and Bit Error Rate (BER). We investigate the efficacy of IPS through a series of controlled, laboratory experiments that consider how the candidate signals respond when fiber is subjected to a range of stresses.   We believe that IPS offers the potential to transform the practice of scientific, commercial and public safety-related vibration monitoring applications by providing a highly-sensitive platform that is available at a global scale. \end{abstract}

\section{Introduction} \label{sec:introduction} The ability to accurately measure deformations and vibrations is central to a variety of public safety, scientific and commercial applications including seismology, non-destructive evaluation, acoustic and flow monitoring, and security. Measuring displacements and vibrations provides insights on issues related to failure potentials, misalignments, utilization, wear and tear, flow rates, etc. that can be used for real-time alerting or post-facto assessment of system behaviors. 

While the notion of monitoring the dynamic response of systems is intuitive, there can be significant challenges gathering measurements depending on the specific application. The key requirements for measuring mechanical responses include precision and accuracy of measurement, scale and range of sensors, time response, and deployability. Standard devices for measuring deformation and vibrations include laser displacement sensors, strain gauges, geophones, MEMS and piezoelectric accelerometers, gyroscopes, etc. Selecting a transducer depends on the requirements for deployment and sensitivity. 

In the past several years, Distributed Strain Sensing (DSS) and Distributed Acoustic Sensing (DAS) have been demonstrated as powerful alternatives to standard methods for displacement and vibration monitoring. DSS and DAS are based on using purpose-built optical hardware, which generates light pulses on an optical fiber strand and then measures the location of Brillouin or Rayleigh scattering events using optical time domain reflectometry (OTDR) to sense strain and strain rates.  These techniques offer highly sensitive monitoring capability, accurate source localization, high sensing density, and the ability to monitor over long distances. DSS and DAS typically operate on dedicated fiber strands up to 50 km long, deployed in a location of interest.  Exploiting dark fiber in existing fiber conduits is an important recent advance in using DAS over long distance~\cite{Franklin19,Wellbrock19}.

One of the main disadvantages of DSS- and DAS-based distributed sensing arrays is the need to install dedicated fiber in the area to be monitored. This problem may be overcome to a certain extent by using dark fiber in existing conduits if it happens to be available in the target area. However, the issue of using expensive, purpose-built hardware remains.  

In this paper, we describe a new method for deformation and vibration measurement and monitoring that we call {\em Internet Photonic Sensing} (IPS). IPS is based on the observation that is central to the aforementioned distributed sensing techniques, namely that the behavior of light transmitted through optical fiber will change when the fiber is exposed to a broad range of vibrations. Our objective in this paper is to take steps toward demonstrating the efficacy of IPS by investigating how stresses on optical fiber that transmits standard Internet data can be identified through simple measurements.   To that end, we report results of a series of laboratory-based experiments that examine how stress on optical fiber affects two simple metrics available from standard Internet transport hardware.
Our vision for IPS is highly sensitive, low-cost, world-wide vibration monitoring capability using already-deployed infrastructure.

\section{Background} \label{sec:background} IPS is motivated by recent advances in {\em (i)} the use of fiber-optics for deformation sensing and {\em (ii)} the wide deployment of coherent optical systems in today's Internet.  In this section, we provide a description of these technologies and the system component model that is a starting point for our work.

{\bf Fiber-optic deformation sensing.} Fiber-optic sensing technology has been applied for over a decade to study deformation in buildings, dams, tunnels, and bridges for ``structural health monitoring" (SHM)~\cite{Glisic07}, and for studying natural earth processes such as earthquakes, landslides and glacier movement. Fiber-optic sensors have many benefits over other strain~\footnote{Strain $\varepsilon$ is a dimensionless metric for deformation caused by a stress on a body.  } sensors including sensitivity, stability, effective range and spatial resolution (the ability to identify strain location).  Fiber-optic sensors can be divided into two categories: {\em Fiber Bragg Grating (FBG)} and {\em distributed systems}.
A FBG sensor is a discrete device that is based on inscribing a pattern in the core of single mode fiber that changes its refractive index. A FBG sensor identifies strain changes from the wavelength peak of the reflected light ~\cite{Kersey97}.
By increasing the acquisition rate of the FBG, sensors can be used to monitor dynamic responses. 

Distributed systems utilize backscatter energy as an indication of strain.  These systems are based on Brillouin Optical Time-Domain Optical Reflectometry (BOTDR~\cite{Sun10,Thevenaz10}), or Coherent Optical Time-Domain Reflectometry (C-OTDR~\cite{Bakku15}).   BOTDR is used in DSS to sense strain and the C-OTDR is used in DAS~\cite{Daley13} to sense strain rates. The DSS sensitivity is on the order of microstrain over distances of tens of kilometers with spatial resolution  on the order of millimeters. DAS senses strain rates with sensing lengths of about 10 m, measurement separation of 1 m, and can be deployed over distances of several kilometers. The associated sampling frequencies for both techniques range from the order of seconds to milliseconds. 


{\bf Coherent optics in the Internet.} In the last ten years, physical layer networking hardware~\footnote{For the remainder of this paper we will refer to these as {\em transport systems and transport hardware}.} has seen a significant shift to coherent optical communications equipment that
employs a combination of modulation techniques to encode data in laser light, often at carrier wavelength of 1550 nm (193.5 THz).  These systems rely on high speed analog-to-digital converters that sample the received signal at a rate commensurate with the bandwidth of the optical channel, allowing full characterization of the received optical signal. 
After conversion, a digital signal processor (DSP) is responsible for carrier frequency recovery, carrier phase recovery, channel equalization and down-sampling at the symbol rate.  

The carrier phase and polarization provide direct paths to estimate cable deformation caused by vibrations.   However, measurements of phase and polarization characteristics are not surfaced through standard network device management interfaces.  Our challenge is that the DSPs in transport systems are designed to remove the variations in received light signals caused by vibrations, which are exactly the signals that we want to use in IPS.  In a very real sense, this ``Internet garbage" is ``distributed strain and vibration sensing gold".

{\bf System component model for transport hardware.}  We investigate the efficacy of IPS based on a simple component model for transport hardware. We assume that stresses on optical fibers will cause changes in transmitted light signals as they propagate through fiber~\cite{Glisic07}.  While the optical signals such as carrier offset, relative phase, and the channel parameters are estimated by the DSP after A-to-D conversion, they are not typically available in higher level interfaces and application program interfaces (API) in commodity transport hardware.  Thus, in this paper we consider how simple metrics that are readily available from transport hardware can reveal when fiber is subjected to vibration stress.  This approach enables experiments on commodity transport hardware and suggests the possibility of wide deployment of IPS if our hypothesis is correct.  

We consider two signals in our study:  received Optical Signal Strength (OSS in units of dBm) and Bit Error Rate (BER - bit errors per second), which are commonly recorded in management information bases (MIBs) on transport hardware and are readily available for measurement via Simple Network Management Protocol (SNMP).  These signals were selected based on our broad consideration of signals available in MIBs that might change when fiber is subjected to vibration stress.  Other signals and combinations of signals will be investigated in future work.

OSS and BER have a straightforward relationship in an \emph{additive white noise Gaussian noise} (AWGN) channel, which is a common starting assumption for noise analysis in a coherent optical channel.  As the received OSS decreases, the signal-to-noise ratio (SNR) of the communications channel decreases proportionally.  This decrease in OSS results in an increased BER.  The BER of a Quadrature Phase Shift Keying (QPSK) communications signal (a modulation technique for encoding bits in coherent optical systems) in an AWGN channel can be derived from the SNR ~\cite{proakis2001digital}:  
$\mathrm{BER} = Q(\sqrt{2 \ \mathrm{SNR}})$
where $Q(\cdot)$ is the q-function\footnote{The q-function can be expressed as $Q(x) = \frac{1}{\sqrt{2\pi}} \int_x^{\infty} e^{\frac{-x^2}{2}} dx$.}.  Similar expressions are available for higher order modulation schemes.

{\bf Related physical layer assessments.} There have been several prior studies of optical signal characteristics measured on live fiber routes~\cite{Brodsky06,Feuerstein05,Woodword14}.  These studies  characterize signal perturbations due to environmental conditions, manufacturing defects and deployment issues. However, we are not aware of prior studies that report on how vibrations change optical signals. Monitoring signals in transport hardware for the purpose of network health assessment has also been the subject of several prior studies~\cite{Ghobadi16,Hasegawa19}.  
These studies highlights the opportunities for insights from studying physical layer behavior.  We take that one step further in proposing the opportunistic use of physical layer signals for vibration monitoring in IPS.


\section{Methodology} \label{sec:methods} Our first step toward assessing the efficacy of IPS was to conducted a series of controlled, laboratory-based experiments using standard transport hardware to measure the cause-effect relationship between stress along a length of the optic fiber and our two candidate data transmission parameters:  OSS and BER.  The sending and receiving transport hardware consisted of transponders (Infinera AOLM-500-T4-1-C6) connected to Bandwidth Multiplexing Modules (BMMs) (Infinera BMM2-4-CX2-MS-A cards in separate DTC-A chassis), which are commonly found in the Internet.  The transceivers in these systems operate in the C-band (1550nm) and were configured based on fiber lengths (thus OSS values for different lengths of fiber used in experiments will be different).  During all of our experiments the transponders constantly send/receive streams of empty Optical Data Units (ODUs) at 100 Gbps (ODUs are normally populated with data from higher layers). The bit streams that make up the ODUs are modulated via Dual Polarization-Quadrature Phase Shift Keying (DP-QPSK).  The BMMs are connected via standard 9/125 single-mode fiber (9 and 125 are the diameters of the fiber's core and cladding respectively in micrometers). 

\begin{figure}[ht!]
\centering
\includegraphics[width=8.5cm]{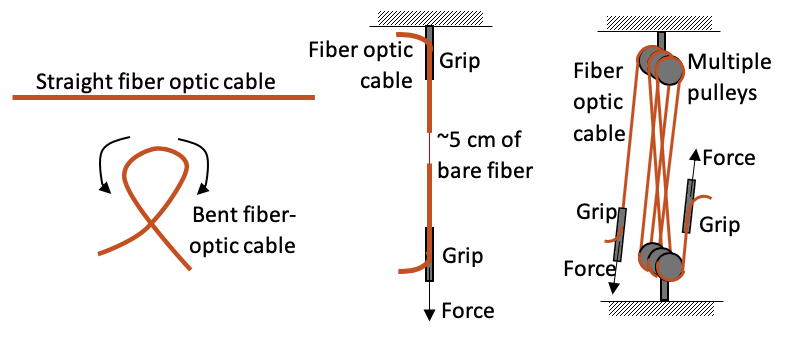}
\caption{Mechanical deformation of the fiber: (a) bending, (b) pullout tension to a short section (1.5 m) of fiber, and (c) pullout tension to a long section (12.1 m) of fiber.}
\label{fig:experiments}
\end{figure}

As shown in Figure~\ref{fig:experiments}, we use three different configurations for our experiments:  static bend tests through different bend radii on a 7 m segment of fiber, pull stress applied to a relatively short segment of fiber (1.5 m) and pull stress applied to a longer length of fiber (12.1 m) enabled through a system of pulleys.  Bending fiber is an extreme form of stress that degrades the signal by causing light to leak through cladding~\cite{Jay10}.  The pull stress testing recognizes the relationship between strain and fiber length~\cite{Henault12}:  longer lengths of fiber may result in greater OSS and BER changes for the same level of stress. 

\begin{figure}[ht!]
\centering
\includegraphics[width=8cm]{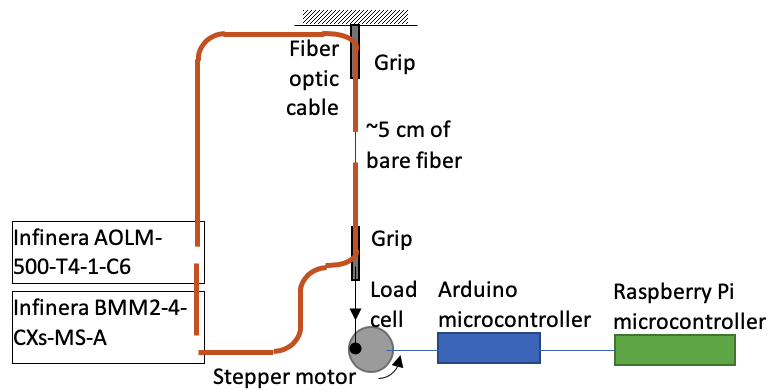}
\caption{Test environment and electronic peripherals for generating pull stress, control and communication.}
\label{fig:setup}
\end{figure}

One of the main challenges in our work was to design a mechanical system to apply stress to an optical fiber.  Ideally, the system would enable us to experiment over a range of stress levels that are consistent with the deformation and vibrations that are the target for DAS.  Such strains typically occur in a range of 100's nanostrains to microstrains~\footnote{1 microstrain amounts to changing the distance between two points that are 1 km apart by 1 mm.}.  

The pull stress system that we built is depicted in Figure~\ref{fig:setup}.  This system was able to subject fiber to tensile force between 50 g and 140 g.  The 50 g stress subjects the fiber to about 100 microstrains~\footnote{Strain $\varepsilon = (F/A)/E$, where F = tension, A = cross sectional area of bare fiber for 250 micrometers diameter, and E is Young's modulus of glass.}.  We argue that the ability to detect cause-effect at this level is indicative of the efficacy of IPS and establishes a critical starting point for future IPS experiments~\footnote{Vibration tables with adjustable frequency and amplitude are commonly used to generate microstrains in fiber for lab-based testing.  We will conduct experiments with vibration tables in future work.}.

Our pull stress system attaches to the cable jacket near a 5 cm portion of exposed fiber via adhesive and using a S-hook attached to a load sensor, which in turn is connected to an Arduino-controlled servo motor (Atmega 328 board). A tensile deformation was applied to the fiber and the reaction force was measured with the HX711 Arduino-controlled load cell amplifier.  The time series of the applied force was collected and transmitted using a Raspberry Pi 2 Model B Rev 1.1. 

Simple Network Management Protocol (SNMP) is used to query the MIBs on the BMMs.  Results of the SNMP queries were stored in files for post processing.  The information maintained in the MIBs is structured in trees, with branches and end nodes referred to as {\em object identifiers} (OIDs).  Some of OIDs we considered include the Number Of Corrected Zeros, Number Of Corrected Ones, Number of Corrected Bits, Number of Corrected Words, Q Factor, Pre Forward Error Correction (PreFEC), Bit Error Rates (BerPreFec), channel chromatic dispersion and channel Q value.  We were able to query the receiving BMM for OSS and BER at an approximate (maximum) rate of 1 query per second.  While this rate is sufficient for our experiments where stress is applied over tens of seconds to minutes, the SNMP query and MIB update rates are important factors for IPS. Ideally, rates should be 20 queries per second to sample a broader range of cultural and natural seismic signals.

\section{Results} \label{sec:results} In this section we report results from our laboratory experiments.  In each case we give representative examples of tests that were repeated multiple times.  We provide perspective on our lab experiments by reporting measurements of BER and OSS from transport hardware deployed in the live Internet, and on an automated method for detecting changes in BER and OSS.

{\bf Bend tests.} Our first experiment examined the impact of a single bend strain on BER and OSS.  We varied the radius of the bend, and separated the bend periods with periods of rest.  Each bend strain period is 2 minutes.

\begin{figure}[ht!]
\centering
\includegraphics[width=.95 \linewidth]{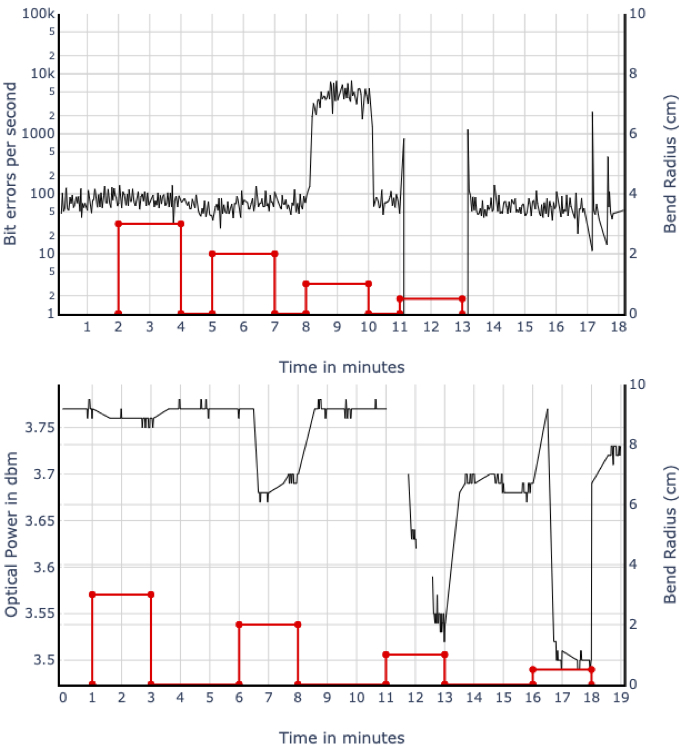}
\caption{BER (above) and OSS (below) response to bend stress with radius of 3 cm,  2 cm, 1 cm and 0.5 cm each for a duration of 2 minutes, with rest periods of 1min (BER) and 3 min (OSS).}
\label{fig:bend}
\end{figure}

As shown in Figure~\ref{fig:bend}, BER with 1 minute rest periods remains constant for bend radius of 3 cm and 2 cm indicating that the DSP and error correction can compensate for changes to light signal characteristics caused by stress in the fiber at this level.  However, when the bend radius reaches 1 cm BER jumps by two orders of magnitude indicating that the changes to light signal characteristics caused by stress in the fiber have resulted in significant degredation of the optical channel. Finally, the figure shows a complete loss of signal during the bend stress period with a radius of 0.5 cm.  This indicates a complete loss of bits from the transmitted signal, thus no BER is measured during this period.

Figure~\ref{fig:bend} also shows that OSS degrades proportionally for all bend radii.  In fact, OSS is sensitive to even light touching of fiber when changing bend radius, which is why we show results with a 3 minute rest period.  While the bend results clearly highlight the cause-effect of stress and the target signals, strains exerted at 1 cm and below are far beyond what would be exerted in the vibration monitoring scenarios that we envision for IPS.

\begin{figure}[ht!]
\centering
\includegraphics[width=\linewidth]{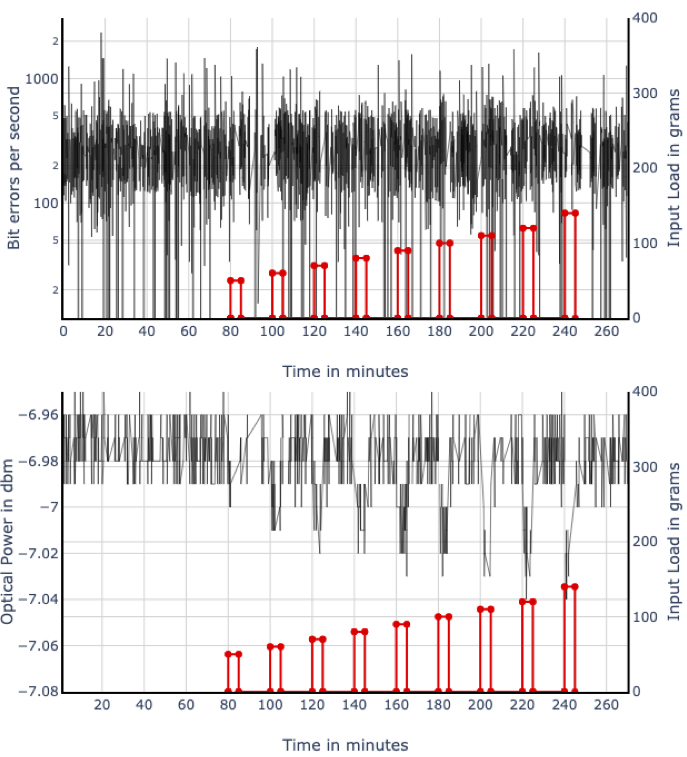}
\caption{BER (above) and OSS (below) response to pull stress between 50 g and 140 g (in 10 g increments) on the long section of fiber. Load Period: 5 min, Rest period: 15 min.}
\label{fig:varyload}
\end{figure}

{\bf Pull stress tests.} We conducted pull stress tests using both short and long segments of fiber as shown in Figure~\ref{fig:experiments}.   We did not find significant differences in measured BER and OSS in the short vs. long fiber experiments, thus we only report the results of the experiments on long fiber segments in this paper.  The focus of our experimental configurations was on {\em (i)} stress level, {\em (ii)} stress duration and {\em (iii)} rest duration between stresses. Our rationale for considered these aspects is based on our system component model approach.  Namely, we want to understand how the target signals react to different stresses recognizing that the DSP is responding to changes in the optical channel. 

Figure~\ref{fig:varyload} shows how BER and OSS react to different stress levels (load applied in increments of 10 g from 50 g to 140 g) where the duration of the stress and rest period between is constant at 5 min and 15 min respectively.  The figure shows that under these stresses, BER is not obviously affected, but OSS is clearly affected even under the 50 g load.

\begin{figure}[ht!]
\centering
\includegraphics[width=\linewidth]{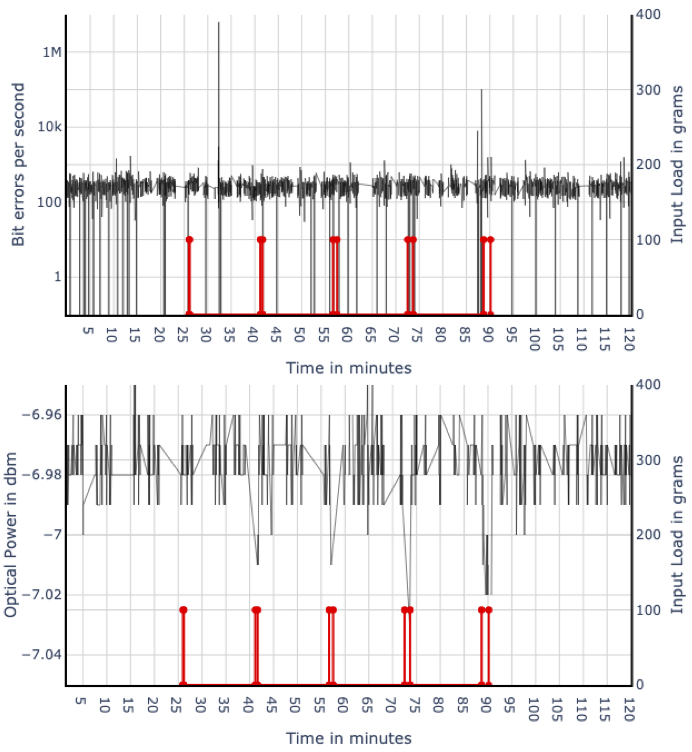}
\caption{BER (above) and OSS (below) response for stress duration between 10 s and 50 s (in 10 s increments), on the long section of fiber. Load: 100 g, Rest period: 15 min.}
\label{fig:varyld}
\end{figure}

Figure~\ref{fig:varyld} shows how BER and OSS react to 100 g stress applied over durations that vary between 10 sec and 50 sec at 10 sec increments with a rest period of 15 min. The figure shows that BER is inconclusive regardless of the duration of the stress.  OSS is not affected by a stress applied for 10 sec, but is clearly reacting to stresses applied for longer durations.

\begin{figure}[ht!]
\centering
\includegraphics[width=\linewidth]{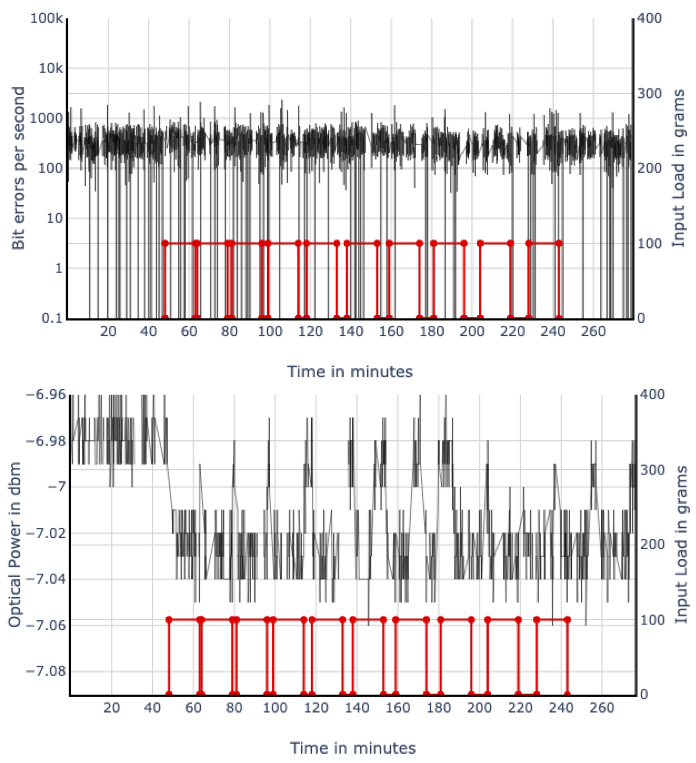}
\caption{BER (above) and OSS (below) response for rest duration between 1 min and 10 min (in 1 min increments) on the long section of fiber. Load: 100 g, Load Period: 15 min.}
\label{fig:varyrd}
\end{figure}

Figure~\ref{fig:varyrd} shows how BER and OSS react to a 100 g load applied over a duration of 15 min with rest periods that vary between 1 min, and 10 min. The figure shows that BER is inconclusive at this level of stress regardless of the duration of the rest period.  OSS returns to the pre-stress level more fully with longer rest durations up to about 7 min. After reaching the maximum rest duration, recovery has to build up all over again.

\begin{figure}[ht!]
\centering
\includegraphics[width=\linewidth]{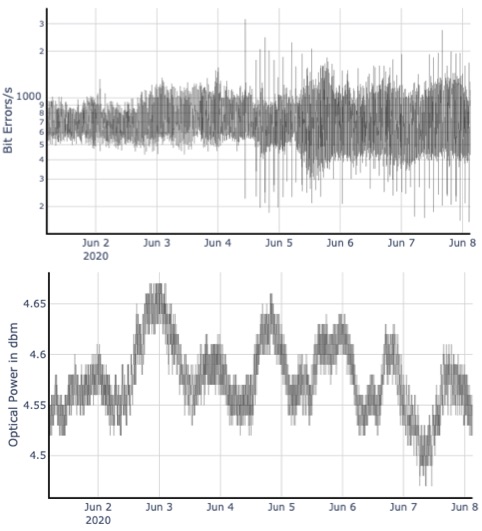}
\caption{BER (above) and OSS (below) measurements taken over a period of 7 days from a 130-mile link on live transport hardware.}
\label{fig:longtermA}
\end{figure}

{\bf BER and OSS measurements on live links.} To provide perspective on our laboratory assessment of cause-effect on BER and OSS, we set up SNMP-based monitoring on two WAN links that span approximately 90 miles Madison WI, and Milwaukee, WI and 130 miles between Madison, WI and Chicago, IL.  We collected signals every 10 seconds over a period of several weeks on both links. Figure~\ref{fig:longtermA} shows a representative snapshot of BER and OSS from one of the links over a 7-day period.  While we cannot establish direct cause-effect due to vibration, we see that OSS follows a diurnal cycle which may reflect, among other things, increased traffic along the end-to-end route, which is known to follow an interstate highway.  BER shows a variety of interesting features, which we are investigating in on-going work.

\begin{figure}[htb]
\includegraphics[width=\linewidth]{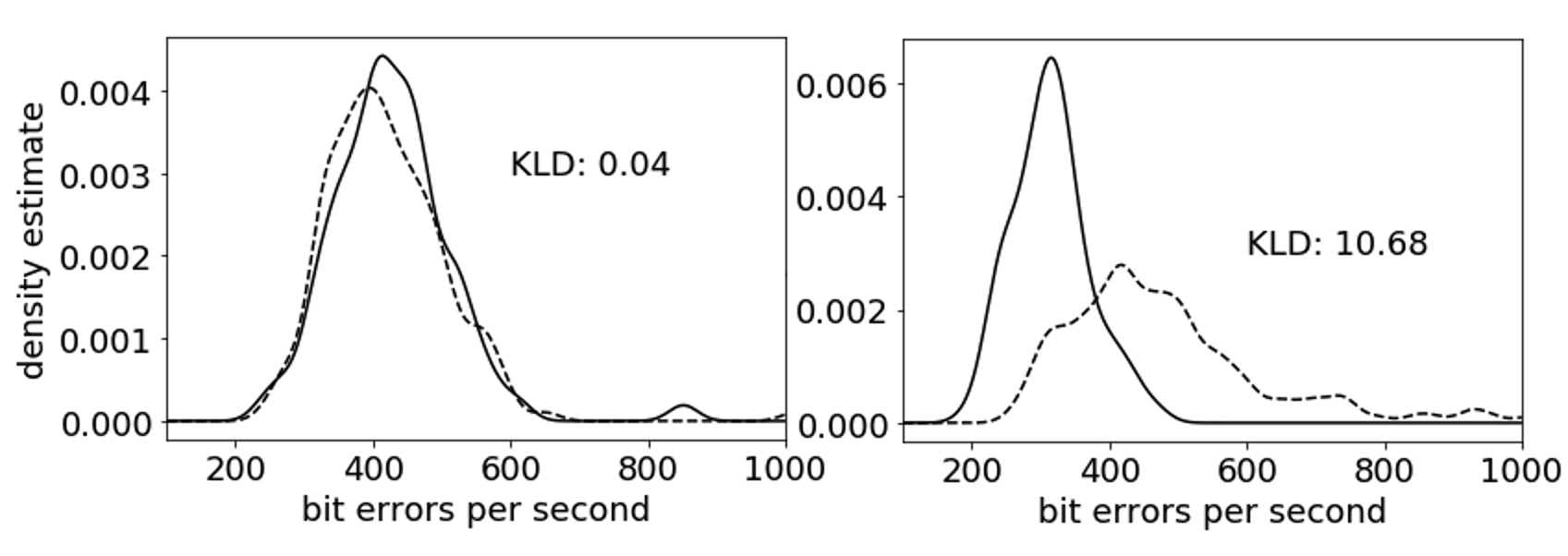}
\caption{Kernel density estimates of bit errors per second when no change is detected (left) and when change is detected (right).  The solid trace corresponds to a density estimate using $200$ samples, while the dashed line corresponds to the density estimate using the subsequent $200$ samples.  Gaussian kernel, bandwidth $= 20$.  The Kullback-Leibler divergence (KLD) is shown on the plot.  }
\vspace{-3mm}
\label{fig:change}
\end{figure}

{\bf Change point detection.} An objective of IPS is the ability to automatically detect changes in signals from many locations that indicate vibrations.  At a high level, this is a change point detection problem~\cite{Tartakovsky14}, which has similarities to other networking studies {\em e.g.,}~\cite{Ahmed08,Wang04}.  To provide perspective on how this might work, we developed a change point detector that estimates the histogram of a signal using a non-overlapping window~\cite{aminikhanghahi2017survey}.  The histograms are estimated using kernel density estimation, and a change is detected when the Kullback-Leibler (KL) divergence between two adjacent density estimates exceeds a threshold.  An example is shown in Figure~\ref{fig:change}.  We applied this detector to BER and OSS traces and confirmed its ability to detect changes with low false detection rates.  A systematic investigation of change point detection for IPS is beyond the scope of this paper, but we plan to investigate this issue in detail in future work.

\vspace{-3mm}
\section{Discussion} \label{sec:discussion} 
\begin{figure}[ht!]
\centering
\includegraphics[width=6.0cm]{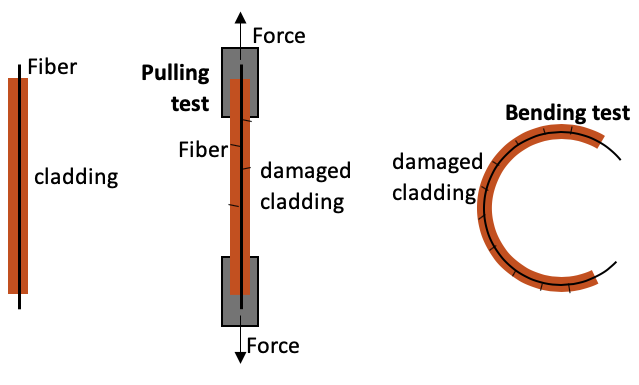}
\vspace{-3mm}
\caption{Potential damage mechanisms of the fiber and cladding.}
\label{fig:damage}
\vspace{-3mm}
\end{figure}

{\bf Weight and bend testing.} The pulley system in our laboratory tests has similarities to work by Henault {\em et al.} who conducted ``pull-out" tests in which marks were placed 3.5 cm apart on 9-cm of bare fiber and the jacket of a 30-cm long sample~\cite{Henault12}.  As tensile force was applied, the separation increased linearly up to about 400 g. For higher loads, multiple shearing events occurred between the outer and inner layers because each concentric layer has different Young's modulus. The behavior can be likened to ``stick-slip" behavior in which minor strength variability between layers eventually leads to different portions of the glass/coating interface to slip. Differential displacement between the jacket and the inner 250-micrometer coating might produce small ``microseisms" due to microcracking as illustrated in Figure~\ref{fig:damage}. The acoustic emission noise could transiently increase BER. 

Shearing between the primary coating and glass cladding during the IPS weight and bend tests similar to that observed by Henault {\em et al.} might be expected. The observed decrease in OSS is hypothesized to be caused by microbending effects of optical leakage or cable damage~\cite{Jay10}. The attenuation coefficient increased approximately linearly for weights between 50 g and 140 g. The signal strength recovered to its unweighted level, given a sufficient resting period.  Full recovery required about 15 minutes; otherwise the recovery was incomplete. The load-recovery cycle became unclear after several cycles when a new recovery cycle began. The bend test showed no optical signal attenuation until the bend radius was reduced to a critical value of 0.5 cm. 

While microbending is generally undesirable in fiber-optic communications~\cite{Berthold99}, the effect has also been the basis for designing strain sensors~\cite{Yao83,Ula17}. One way to test the hypothesis that optical signal attenuation resulted from microbending damage is to use fiber stripped of its jacket.  We are investigating these effects in greater detail in on-going work.

{\bf Phase changes and strain.} Our laboratory weight-and-bend tests showed high correlation with OSS but much less so with BER. We hypothesized that these results could be explained by relatively high shear stresses between the primary coating and the cladding, which created microbends and hence optical attenuation. We turn now to speculate how much smaller, elastic longitudinal strain might be observed using IPS. The physics underlying this IPS application is that strain in optical fiber changes the index of refraction through the photoelastic effect. We first draw attention to how this basic principle presents similarities between IPS and DAS. DAS can detect changes on the order of 100 nanostrain over distances of a few meters. A DAS interrogator sends 1550 nm laser pulses separated by 100 microseconds and analyzes the Rayleigh backscattered light for the phase change of two locations a few meters apart ($\Delta z$).  The change in strain ($\Delta \varepsilon$) is 
$\Delta \varepsilon = K \frac{\Delta \Phi}{\Delta z} $ where K is a constant that depends on the wavelength of the laser, the index of refraction, and the photoelastic constant~\cite{Chen17}.

This photoelastic proportionality applies to network data transmission. Even the same C-band laser frequency of 1550 nm is typical for which $K = 100n \varepsilon \ /\mathrm{rad}$.  The key conceptual link to IPS is to use BER during modulated coherent optical data transmission as a measure of phase change, $\Delta \Phi$, over a transmission distance, $\Delta z$, between transceivers. The photoelastic proportionality shows that the phase change over a path distance $\Delta z$ is the product $\Delta \varepsilon  \Delta z$. We define a critical phase change, $\Delta \Phi_{crit}$, large enough to produce statistically measurable BER changes. In the case of QPSK, $\Delta \Phi_{crit} \approx \frac{\pi}{4}$.  This condition is met easily as 100 nanostrain over 100 m ({\em i.e.,} the approximate separation between transceivers in an university campus network) gives a phase change of 100 rad, requiring phase unwrapping (estimation of phase at a sufficiently high rate to unwrap the phase).  DSPs in coherent optical receivers estimate, track, and correct the relative laser phase at rates on the order of microseconds.  The phase estimates would be greatly beneficial to IPS and our efforts, but in general, are not surfaced in networking management interfaces.  In this way, the DSP masks the vibration induced signals that would otherwise be more visible in BER and OSS signals studied herein.  

 A number of possible future experiments are envisaged at the laboratory scale of meters to tens of meters, as well as in local area networks. These include direct measurement of the relative carrier phase using a coherent optical modulation analyzer.  Higher order modulation schemes, such as M-ary Phase Shift Keying techniques could be investigated.  One would anticipate greater BER sensitivity to encoding methods that produce constellation points closely spaced in phase. Distinguishing between OSS and BER effects could be investigated by comparing 16-QAM with QPSK. Comparison of two-way data transmission over the same fiber link could serve to reduce spurious ISP interpretations.  The long-term goal would be to establish the feasibility of using network data communication systems for ground motion monitoring in urban settings where transceivers are closely spaced.

\vspace{-3mm}
\section{Summary} \label{sec:summary} In this paper we introduce Internet Photonic Sensing -- the use of signals that are available from deployed Internet fiber optic communication hardware for deformation and vibration measurement.  IPS offers the tantalizing possibility of utilizing the gigantic, world wide infrastructure that underpins the Internet for seismologic, infrastructure, security and even communication network monitoring that normally requires deployment of dedicated sensor arrays.  We take first steps toward investigating the efficacy of IPS by conducting a series of laboratory-based experiments that assesses the cause-effect relationship between strain applied to fiber and two standard signals available from transport hardware:  OSS and BER.  We built an apparatus that includes standard Internet transport hardware and enabled us to subject different lengths of fiber to varying strains for varying durations.  
Our system-component analysis of coherent optical data transmission leads to the expectation that OSS and BER will react over several orders of magnitude of combinations of strain amplitude and strained length of fiber with OSS providing the clearest perspective on strains.  Our results point directly to the next steps that need to be taken to realize IPS.   In particular, we plan to {\em (i)} expand our testing infrastructure to enable cause-effect experiments at strain levels that are consistent with a range of standard applications,  {\em (ii)} develop a vibration monitoring system based on an expanded set of received (post-DSP) signals with sensitivity that is sufficient for standard applications, and {\em (iii)} deploy and test our monitoring system in situ.  While this represents a significant undertaking, our analysis shows that it is feasible, and we expect that there are opportunities for further contributions.

\bibliographystyle{IEEEtran} 
\bibliography{paper}

\end{document}